\documentstyle[bo99,epsfig]{article}
\def\ga{\;\rlap{\lower 2.5pt\hbox{$\sim$}}\raise 1.5pt\hbox{$>$}\;}
\def\la{\;\rlap{\lower 2.5pt\hbox{$\sim$}}\raise 1.5pt\hbox{$<$}\;}
\def\gsim{\;\rlap{\lower 2.5pt\hbox{$\sim$}}\raise 1.5pt\hbox{$>$}\;}
\def\lsim{\;\rlap{\lower 2.5pt\hbox{$\sim$}}\raise 1.5pt\hbox{$<$}\;}
\def\lya{Ly$\alpha$}

\title{Probing the Cosmic Dark Age in X--rays}

\author{Z.~Haiman$^{1,2}$}

\affil{1) Department of Astrophysics, Princeton University, 
Princeton, NJ 08544, USA \\
2) Fermi National Accelerator Laboratory,  P.O. Box 500,
Batavia, IL 60510, USA}

\begin{document}

\maketitle

\begin{abstract}
Empirical studies of the first generation of stars and quasars will likely
become feasible within the next decade in several different wavelength bands.
Microwave anisotropy experiments, such as MAP or Planck, will set constraints
on the ionization history of the intergalactic medium due to these sources.  In
the infrared, the Next Generation Space Telescope (NGST) will directly detect
sub--galactic objects at $z\ga 10$.  In the optical, data from the Hubble Deep
Field (HDF) already places a constraint on the abundance of high--redshift
quasars.  However, the epoch of the first quasars might be first probed in
X--ray bands, by instruments such as the Chandra X-ray Observatory (CXO) and
the X-ray Multi-mirror Mission (XMM).  In a 500 Ksec integration, CXO reaches a
sensitivity of $\sim 2\times 10^{-16}~{\rm erg~s^{-1}~cm^{-2}}$. Based on
simple hierarchical CDM models, we find that at this flux threshold
$\sim10^{2}$ quasars might be detectable from redshifts $z\ga 5$, and $\sim 1$
quasar at $z\sim 10$, in each $17^\prime\times17^\prime$ field.  Measurement of
the power spectrum of the unresolved soft X--ray background will further
constrain models of faint, high--redshift quasars.  \keywords{cosmology:
theory; quasars: general; black hole physics}
\end{abstract}

\section{Introduction}

The cosmic dark age ended when the first gas clouds condensed out of the
primordial fluctuations at redshifts $z=10-20$ (Peacock 1992; Rees 1996).
These condensations are likely the sites where the first clusters of stars and
the first quasar black holes appeared, giving birth to the first
``mini--galaxies'' or ``mini--quasars'' in the Universe.  Despite the lack of
observational data, this epoch has become a subject of intense theoretical
study in the past few years.  The recent interest can be attributed to
forthcoming instruments: {\it NGST} could directly image sub--galactic
objects at $z\ga 10$ in the infrared, while microwave satellites such as MAP or
Planck could measure signatures from the reionization of the intergalactic
medium (IGM).  Currently, bright quasars are detected out to $z\sim 5$ (Fan et
al. 1999).  Although the abundance of optically and radio bright quasars
declines at $z\gsim 2.5$ (Schmidt et al. 1995; Shaver et al. 1996), a recent
determination of the X--ray luminosity function (LF) of quasars from ROSAT data
(Miyaji et al. 1998a) has not confirmed this decline. In this contribution, we
point out that future X--ray observations might provide yet another probe of
the first quasars and the end of the dark age at $z\sim10$, and that X--ray
data might be uniquely useful in distinguishing quasars from stellar systems.

\section{The Appearance of the First Quasars and Stars}

In popular Cold dark matter (CDM) cosmologies,the first baryonic objects appear
near the Jeans mass ($\sim 10^6~{\rm M_\odot}$) at redshifts as high as
$z\sim30$ (Haiman \& Loeb 1999b, and references therein).  At any redshift, the
mass function of collapsed dark halos is given to within a factor of two by the
Press--Schechter formalism.  Following collapse, the gas in the first baryonic
condensations is virialized by a strong shock (Bertschinger 1985). Provided it
is able to cool on a timescale shorter than the Hubble time, the shock--heated
gas continues to contract. Depending on the details of the cooling and angular
momentum transport, the gas then either fragments into stars, or forms a
central black hole exhibiting quasar activity.  Although the actual
fragmentation process is likely to be rather complex, the average fraction
$f_{\rm star}$ of the collapsed gas converted into stars can be calibrated
empirically so as to reproduce the average metallicity observed in the Universe
at $z\approx 3$.  The observed ratio, inferred from CIV absorption lines in
\lya~forest clouds, is between $10^{-3}$ and $10^{-2}$ of the solar value
(Songaila 1997 and references therein).  If the carbon produced in the early
mini--galaxies is uniformly mixed with the rest of the baryons in the Universe,
this implies $f_{\rm star}\approx$2--20\% for a Scalo stellar mass function.

An even smaller fraction of the cooling gas might condense at the center of the
potential well of each cloud and form a massive black hole, exhibiting quasar
activity.  In the simplest scenario, the peak luminosity of each black hole is
proportional to its mass, and the light--curve, expressed in Eddington units,
is a universal function of time.  Indeed, for a sufficiently high fueling rate,
quasars are likely to shine at their maximum possible luminosity, which is some
constant fraction of the Eddington limit, for a time which is dictated by their
final mass and radiative efficiency.  Here we assume that the black hole mass
fraction $r\equiv M_{\rm bh}/M_{\rm halo}$ obeys a log-Gaussian probability
distribution, $p(r)=\exp[-(\log r - \log r_0)^2/2\sigma^2]$, with $\log
r_0=-3.5$ and $\sigma=0.5$ (Haiman \& Loeb 1999b). These values roughly reflect
the distribution of black hole to bulge mass ratios found in a sample of 36
local galaxies (Magorrian et al. 1998) for a baryonic mass fraction of $\sim
(\Omega_{\rm b}/\Omega_0)\approx 0.1$.  We further postulate that each black
hole emits a time--dependent bolometric luminosity in proportion to its mass,
$L_{\rm q}\equiv M_{\rm bh}f_{\rm q}=M_{\rm bh}L_{\rm Edd} \exp(-t/t_0)$, where
$L_{\rm Edd}=1.5\times10^{38}~M_{\rm bh}/{\rm M_\odot}~{\rm erg~s^{-1}}$ is the
Eddington luminosity, $t$ is the time elapsed since the formation of the black
hole, and $t_0=10^6$ yr is the characteristic quasar lifetime.

\begin{figure}[t]
\centerline{\psfig{file=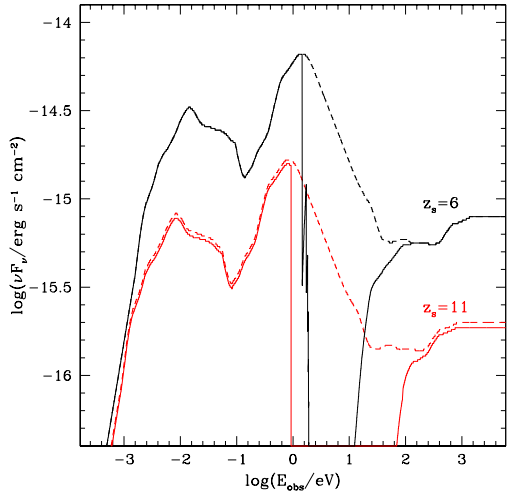, width=7cm}}
\caption[Spectra] {\it
\label{fig:spectrum} \footnotesize The observed spectra of
quasars with a central black hole mass of $M_{\rm bh}=10^8~{\rm M_\odot}$.  The
upper curves correspond to a source redshift of $z_{\rm s}=6$ and the lower
curves to a source redshift of $z_{\rm s}=11$.  In both cases we assume sudden
reionization at $z_{\rm r}=10$.  The dashed curves show the assumed intrinsic
spectral shape, and the solid curves show the spectra after absorption by
neutral intergalactic H and He.}
\end{figure}

Finally, we assume that the shape of the emitted spectrum follows the mean
spectrum of known quasars (Elvis et al. 1994) up to a photon energy of 10 keV.
We extrapolate the spectrum up to $\sim 50$ keV, assuming a spectral slope of
$\alpha$=0 (or a photon index of -1).  For reference, Figure~1 shows the
adopted spectrum of quasars, assuming a black hole mass $M_{\rm bh}=10^8{\rm
M_\odot}$, placed at two different redshifts, $z_{\rm s}=11$ and $z_{\rm s}=6$,
and processed through the IGM, and assumed that reionization occurred at
$z_{\rm r}=10$ and that at higher redshifts the IGM was homogeneous and fully
neutral. At lower redshifts, $0<z<z_{\rm r}$, we included the hydrogen opacity
of the Ly$\alpha$ forest given by Madau (1995), extrapolating his fitting
formulae for the evolution of the number density of absorbers beyond $z=5$ when
necessary.  As Figure~1 shows, the minimum black hole mass detectable by the
$\sim 2\times 10^{-16}~{\rm erg~s^{-1}~cm^{-2}}$ flux limit of {\it CXO} (see
below) is $M_{\rm bh}\sim 10^8~{\rm M_\odot}$ at $z=10$ and $M_{\rm bh}\sim
2\times10^7~{\rm M_\odot}$ at $z=5$.  In our model, the corresponding halo
masses are $M_{\rm halo}\sim 3\times10^{11}~{\rm M_\odot}$, and $M_{\rm
halo}\sim 6\times 10^{10}~{\rm M_\odot}$, respectively.  Although such massive
halos are rare, their abundance is detectable in wide-field surveys.

\begin{figure}[t]
\centerline{\psfig{file=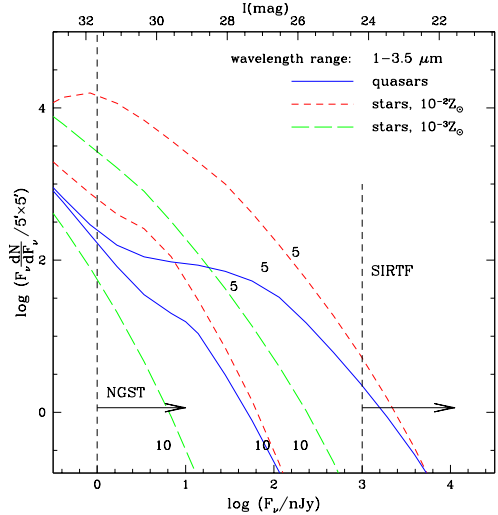, width=7cm}}
\caption{ {\it \label{fig:ncounts}\footnotesize Infrared Number Counts. The
solid curves refer to quasars, while the long/short dashed curves correspond to
galaxies with low/high normalization for the star formation efficiency.
The curves labeled ``5'' or ``10'' show the cumulative number of objects with
redshifts above $z=5$ or 10.}}
\end{figure}

\section{Infrared: Expected Counts with NGST}

The Next Generation Space Telescope\footnote{see http://www.ngst.nasa.gov}
({\it NGST}) will be able to detect the early population of mini--galaxies and
mini--quasars.  {\it NGST} is scheduled for launch in 2008, and is expected to
reach an imaging sensitivity of $\sim 1$ nJy (S/N=10 at spectral resolution
$\lambda/\Delta\lambda=3$) for extended sources after several hours of
integration in the wavelength range of 1--3.5$\mu$m. Figure~2 shows the
predicted number counts in the mini--galaxy and mini--quasar models described
above, in a $\Lambda$CDM cosmology with ($\Omega_0,\Omega_\Lambda, \Omega_{\rm
b},h,\sigma_{8h^{-1}},n$)=(0.35, 0.65, 0.04, 0.65, 0.87, 0.96), normalized to a
$5^{\prime}\times5^{\prime}$ field of view.  This figure shows separately the
number per logarithmic flux interval of all objects with redshifts $z>5$ (thin
lines), and $z>10$ (thick lines). As the figure shows, {\it NGST} will be able
to probe about $\sim100$ quasars at $z>10$, and $\sim200$ quasars at $z>5$ per
field of view.  The bright--end tail of the number counts approximately follows
a power law, with $dN/dF_\nu\propto F_\nu^{-2.5}$.  The dashed lines show the
corresponding number counts of mini--galaxies, assuming that each halo
undergoes a starburst that converts a fraction of 2\% (long--dashed) or 20\%
(short--dashed) of the gas into stars.  These lines indicate that {\it NGST}
would detect $\sim40-300$ mini--galaxies at $z>10$ per field of view, and
$\sim600-10^4$ mini--galaxies at $z>5$.  Unlike quasars, galaxies could in
principle be resolved if they extend over a scale comparable to the virial
radius of their dark matter halos (Haiman \& Loeb 1997; Barkana and Loeb 1999).
The supernovae and $\gamma$-ray bursts in these galaxies might outshine their
hosts and may also be directly observable (Miralda-Escud\'e \& Rees 1997).
Finally, we note that recent data in the $J$ and $H$ infrared bands from deep
NICMOS observations of the HDF (Thompson et al. 1999) could already be useful
to constrain mini--quasar and mini--galaxy models.

\begin{figure}[t]
\centerline{\psfig{file=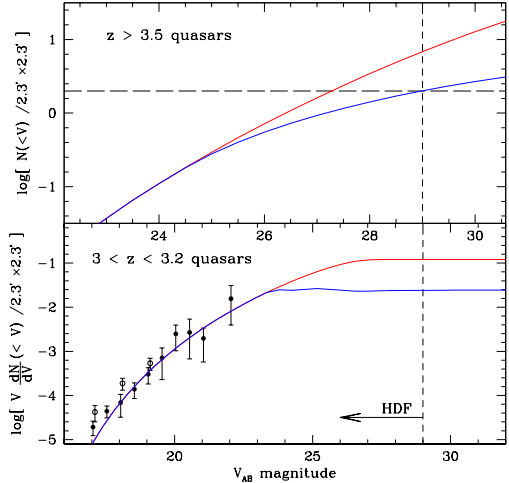, width=7cm}}
\caption{ {\it \label{fig:nicmos}\footnotesize $V$-band number counts for
high--redshift quasars. The lower panel shows differential counts for
$3<z<3.2$, as well as data adapted from Pei (1995).  The upper panel shows
cumulative number counts for $z>3.5$.  The top curve shows predictions of our
standard $\Lambda$CDM model with ${\rm M_{bh}\propto M_{halo}}$.  The bottom
curve shows the counts in a model with halo circular velocities restricted to
$v_{\rm circ} \geq 75~{\rm km~s^{-1}}$.  This model is consistent with the lack
of detections in the HDF.}}
\end{figure}

\section{Optical: Constraints from the Hubble Deep Field}

Although the infrared wavelengths are the best suited to detect the redshifted
UV--emission from objects at $z\sim10$, present data in the optical already
yields a constraint on quasar models of the type described above.  The
properties of faint {\it extended} sources found in the HDF (Madau et al. 1996)
agree with detailed semi--analytic models of galaxy formation (Baugh et
al. 1998).  On the other hand, the HDF has revealed only a handful of faint
{\it unresolved} sources, but none with the colors expected for high redshift
quasars (Conti et al. 1999).  The simplest mini--quasar model described above
predicts the existence of $\sim10$ B--band ``dropouts'' in the HDF,
inconsistently with the lack of detection of such dropouts up to the $\sim50\%$
completeness limit at $V\approx 29$ in the HDF.  To reconcile the models with
the data, a mechanism is needed for suppressing the formation of quasars in
halos with circular velocities $v_{\rm circ} \lsim 50-75~{\rm km~s^{-1}}$ (see
Figure~3 for the counts).  This suppression naturally arises due to the
photo-ionization heating of the intergalactic gas by the UV background after
reionization (Thoul \& Weinberg 1996; Navarro \& Steinmetz 1997).  Alternative
effects could help reduce the quasar number counts, such as a change in the
background cosmology, a shift in the ``big blue bump'' component of the quasar
spectrum to higher energies due to the lower black hole masses in
mini--quasars, or a nonlinear black hole to halo mass relation; however, these
effects are too small to account for the lack of detections in the HDF (Haiman,
Madau \& Loeb 1999).

\begin{figure}[t]
\centerline{\psfig{file=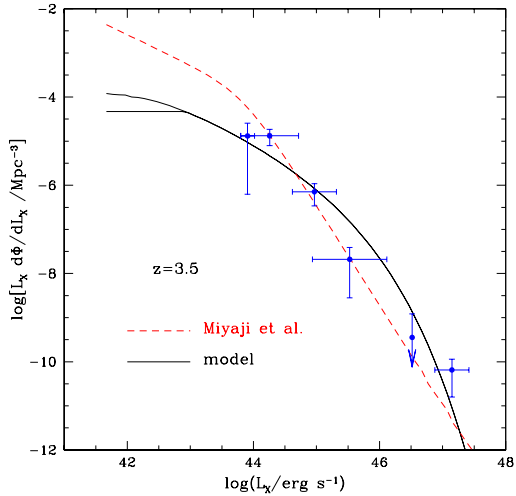, width=7cm}}
\caption[Luminosity Function at z=3.5] {\it
\label{fig:LF} \footnotesize The
predicted X--ray luminosity function at $z=3.5$ in our model (solid curves).
The lower curve shows the effect of a cutoff in circular velocity for the host
halos of $v_{\rm circ}\geq 50~{\rm km~s^{-1}}$. The ROSAT data points are
adopted from Miyaji et al. (1998a) and the dashed curve shows their fitting
formula (for our background cosmology).}
\end{figure}

\begin{figure}[t]
\centerline{\psfig{file=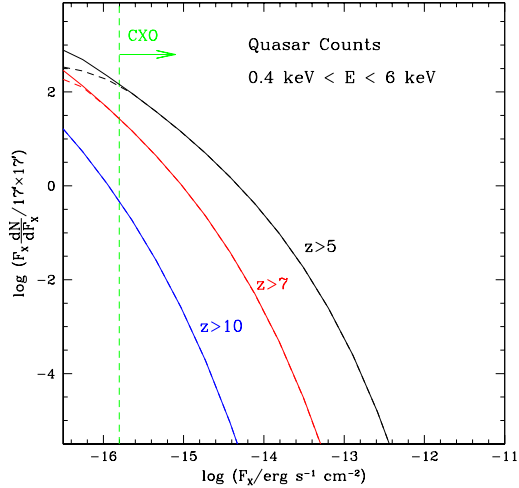, width=7cm}}
\caption[Number Counts] {\it
\label{fig:counts} \footnotesize The total number
of quasars with redshift exceeding $z=5$, $z=7$, and $z=10$ are shown as a
function of observed X-ray flux in the {\it CXO} detection band.  The solid
curves correspond to a cutoff in circular velocity for the host halos of
$v_{\rm circ}\geq 50~{\rm km~s^{-1}}$, the dashed curves to a cutoff of
$v_{\rm circ}\geq 100~{\rm km~s^{-1}}$.  The vertical dashed line show the
{\it CXO} sensitivity for a 5$\sigma$ detection of a point source in an
integration time of $5\times10^5$ seconds.}
\end{figure}

\section{X--Rays: Predictions for CXO}

Quasars can be best distinguished from star forming galaxies at high redshifts
by their X-ray emission. Detections of high-$z$ quasars would therefore be
highly valuable: detections, or upper limits would help in answering the
important question of whether the IGM at $z\lsim 6$ was reionized by stars or
quasars, by yielding constraints on the ionizing photon rate from high--$z$
quasars. The simple quasar--model described above was constructed to accurately
reproduce the evolution of the optical luminosity function in the B--band (Pei
1995) at redshifts $z\gsim 2.2$ (Haiman \& Loeb 1998). However, it yields good
agreement with the data on the X--ray LF, as demonstrated in Haiman \& Loeb
(1999c), and shown here in Figure~4.  We regard this model as a minimal toy
model which successfully reproduces the existing data, and use a
straightforward extrapolation of this model to predict the X--ray number
counts.  In Figure~5, we show the predicted counts in the 0.4--6keV energy band
of the CCD Imaging Spectrometer (ACIS) of {\it CXO}.  Note that these curves
are insensitive to our extrapolation of the template spectrum beyond 10
keV. The figure is normalized to the $17^\prime \times 17^\prime$ field of view
of the imaging chips.  The solid curves show that of order a hundred quasars
with $z>5$ are expected per field at the {\it CXO} sensitivity of $\sim2\times
10^{-16}~{\rm erg~s^{-1}~cm^{-2}}$ for a 5$\sigma$ detection of a point source.
Note that {\it CXO}'s arcsecond resolution will ease the separation of these
point sources from background noise.  The abundance of quasars at higher
redshifts declines rapidly; however, a few objects per field are still
detectable at $z\sim 8$.  The dashed lines show the results for a minimum
circular velocity of the host halos of $v_{\rm circ}\geq 100~{\rm km~s^{-1}}$,
and imply that the model predictions for the {\it CXO} satellite are not
sensitive to such a change in the host velocity cutoff.  This is because the
halos shining at the {\it CXO} detection threshold are relatively massive,
$M_{\rm halo}\sim 10^{11}~{\rm M_\odot}$, and possess a circular velocity above
the cutoff. In principle, the number of predicted sources would be lower if we
had assumed a steeper spectral slope.  For example, as figure~6 shows below,
our model falls short of predicting the hard X--ray background, by about an
order of magnitude at 10 keV.  The difference could be explained by a change in
our template spectrum to include a population of quasars with hard, but highly
absorbed spectra (caused by the denser, and more gas rich hosts at high
redshift).  We note, however, that the agreement between the LF predicted by
our model at $z\approx 3.5$ and that inferred from ROSAT observations would be
upset by such a change, and require a modification of the model that would in
turn tend to counter-balance the decrease in the predicted counts.

\begin{figure}[t]
\centerline{\psfig{file=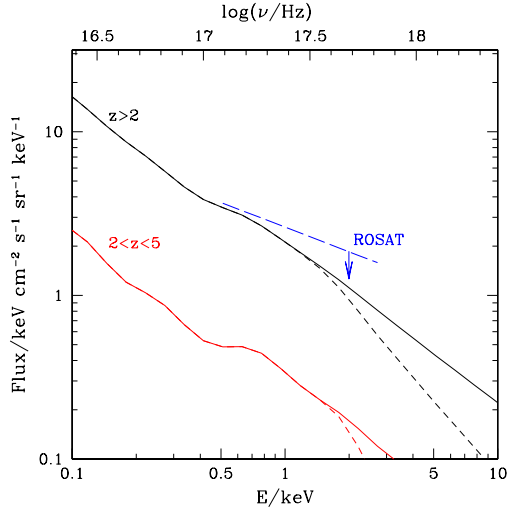, width=7cm}}
\caption[X--Ray Background] {\it
\label{fig:xrb} \footnotesize Spectrum of the 
unresolved soft X--ray background in our model.  We assume that the median
X-ray spectrum of each source follows the mean spectrum of Elvis et al. (1994)
up to 10 keV, and has a spectral slope of 0.5 (solid lines) or -0.5
(short--dashed lines) at higher photon energies.  The lower curves show the
spectra resulting from quasars with redshifts between $2<z<5$, and the upper
curves include contributions from all redshifts $z>2$.  The long--dashed line
shows the unresolved fraction (assumed to be 25\%) of the soft X--ray
background spectrum from Miyaji et al. (1998b).}
\end{figure}

\section{The X--ray Background}

Existing estimates of the X--ray background (XRB) provide another useful check
on our quasar model.  Figure~6 shows the predicted spectrum of the XRB in our
model at $z=0$ (solid lines).  The unresolved background flux is shown,
obtained by summing the emission of all quasars whose individual observed flux
at $z=0$ is below the ROSAT PSPC detection limit for discrete sources of
$2\times10^{-15}~{\rm erg~cm^{-2}~s^{-1}}$ (Hasinger \& Zamorani 1997).  The
short dashed lines show the predicted fluxes assuming a steeper spectral slope
beyond 10 keV ($\alpha=-0.5$, or a photon index of -1.5). The long dashed line
shows the 25\% unresolved fraction of the soft XRB observed with ROSAT (Miyaji
et al. 1998b; Fabian \& Barcons 1992).  This fraction represents the
observational upper limit on the component of the soft XRB that could in
principle arise from high-redshift quasars.  As the figure shows, our quasar
model predicts an unresolved flux just below this limit in the 0.5-3 keV range.
The model also predicts that most ($\gsim 90\%$) of this yet unresolved
fraction arises from quasars beyond $z=5$.  The power spectrum of the
unresolved background therefore might carry information on quasars at $z>5$,
and be useful in constraining the models (Haiman \& Hui 1999, in preparation).
The correlations in the background have recently been measured by Soltan et
al. (1999, see also this Proceedings).

\section{Discussion}

We have demonstrated that state--of--the--art X-ray observations could yield
more stringent constraints on quasar models than currently available from the
Hubble Deep Field (Haiman, Madau, \& Loeb 1999).  The X--ray data might provide
the first probe of the earliest quasars, complementing subsequent infrared and
microwave observations.  More specifically, we have found that forthcoming
X--ray observations with the {\it CXO} satellite might detect of order a
hundred quasars per field of view in the redshift interval $5\la z\la 10$.  Our
numerical estimates are based on the simplest toy model for quasar formation in
a hierarchical CDM cosmology, that satisfies all the current observational
constraints on the optical and X-ray luminosity functions of quasars.  Although
a more detailed analysis is needed in order to assess the modeling
uncertainties in our predictions, the importance of related observational
programs with {\it CXO} is evident already from the present analysis.  Other
future instruments, such as the HRC or the ACIS-S cameras on {\it CXO}, or the
EPIC camera on {\it XMM}, which has a collective area 3--10 times larger than
that of {\it CXO}, will also be useful in searching for high--redshift quasars.

The relation between the black hole and halo masses may be more complicated
than linear.  With the introduction of additional free parameters, a
non--linear (mass and redshift dependent) relation between the black--hole and
halo masses can also lead to acceptable fits (Haehnelt et al. 1998) of the
observed quasar LF near $z\sim3$.  Such fits, when extrapolated to higher
redshift, can result in different predictions for the abundance of
high--redshift quasars.  From the point of view of selecting between these
alternative models, even a non--detection by CXO would be invaluable.  It is
hoped further that either observations of the clustering properties of $z\sim
3$ quasars in the Sloan Digital Sky Survey, or a measurement of the power
spectrum of the soft X--ray background, would break model degeneracies (Haiman
\& Hui 1999, in preparation).

Quasars emit a broad spectrum which extends into the UV and includes strong
emission lines, such as Ly$\alpha$.  For quasars near the {\it CXO} detection
threshold, the fluxes at $\sim 1\mu$m are expected to be relatively high, $\sim
0.5$--$0.8~\mu$Jy. Therefore, infrared spectroscopy of X--ray selected quasars
with the Space Infrared Telescope Facility {\it (SIRTF)} or {\it NGST} can
identify the redshifts of the faint X--ray point-like sources detected by the
{\it CXO} satellite.  Such an approach could prove to be a particularly useful
approach for unraveling the reionization history of the intergalactic medium at
$z\ga 5$.  At present, the best constraints on hierarchical models of the
formation and evolution of quasars originate from the Hubble Deep Field.
However, {\it HST} observations are only sensitive to a limiting magnitude of
$V\sim29$, and cannot probe the earliest quasars, beyond $z\sim 6$.  The
combination of X-ray data from the {\it CXO} satellite and infrared
spectroscopy from {\it SIRTF} and {\it NGST} could potentially resolve one of
the most important open questions about the thermal history of the Universe,
namely whether the intergalactic medium was reionized by stars or by accreting
black holes.

\begin{acknowledgements}
I thank A. Loeb for his advice and guidance throughout many projects, M. Rees
and P. Madau for many useful discussions, and N. White for the invitation to
this stimulating conference. Support for this work was provided by the DOE and
NASA through grant NAG 5-7092 at Fermilab, and a Hubble Fellowship, awarded by
the Space Telescope Science Institute, which is operated by the Association of
Universities for Research in Astronomy for NASA under contract NAS 5-26555.

\end{acknowledgements}

\end{document}